\documentclass[epj,twocolumn]{webofc}
\usepackage[varg]{txfonts}   
\usepackage{multirow,epsfig}
\woctitle{}
\begin{document}
\title{Sensitivity of the LHC to Kaluza-Klein gluon in two b-jets decay channel}
%
%

\author{Masato Arai\inst{1,2}\fnsep\thanks{\email{masato.arai@fukushima-nct.ac.jp}} \and
        Gi-Chol Cho\inst{3}\fnsep\thanks{\email{cho.gichol@ocha.ac.jp}} \and
        Karel Smolek\inst{1}\fnsep\thanks{\email{karel.smolek@utef.cvut.cz}} \and
        Kyoko Yoneyama\inst{3,4}\fnsep\thanks{\email{hep.phys.ocha.ac.jp}}
}

\institute{Institute of Experimental and Applied Physics, Czech Technical University in Prague, Horsk\'a 3a/22, Prague 2, Czech Republic
\and
           Fukushima National College of Technology, Fukushima, 970-8034, Japan
\and
           Department of Physics, Ochanomizu University, Tokyo 112-8610, Japan
\and
           Department of Physics, Bergische Universit\"at Wuppertal Gaussstasse 20, D-42119 Wuppertal, Germany
          }

\abstract{%
We study a possibility of observation of the first Kaluza-Klein (KK) excitation of gluon in a warped extra dimension model at the LHC. In our analysis, we adopt the KK gluon mass and the b-quark coupling to the KK gluon as model parameters and study the sensitivity of the ATLAS experiment to observe the KK gluon through the two $b$-jets channel.
}
\maketitle
\section{Introduction}
\label{intro}

Although the Standard Model (SM) of particle physics has shown a good agreement with almost all data of high energy experiments, we expect new physics beyond the SM from some theoretical motivations such as a gauge hierarchy problem. A warped extra-dimension model proposed by Randall and Sundrum (RS)~\cite{RS} is one of the promising candidates to explain the hierarchy between the Fermi scale and the Planck scale. We suppose the extension of the original RS model, where SM gluons can propagate into the bulk. In such a model, there are Kaluza-Klein (KK) excitations of gluons.

We study a possibility of observation of the first KK excitation of gluon at the LHC. In our analysis, we adopt the KK gluon mass and the $b$-quark coupling to the KK gluon as model parameters and study the sensitivity of the ATLAS experiment to observe the KK gluon through the two $b$-jets channel. More detailed description and analysis is published in our paper \cite{OurPaper}.

\section{Randall-Sundrum model}
\label{sec-2}

In the RS model, there are two 3-branes which are located in different positions in an extra dimension. One of the 3-branes is called a ``visible brane'' in which the SM particles are confined while the other is called a ``hidden brane''. A graviton is allowed to propagate between two branes. With this set-up, the Higgs boson mass can be electroweak scale naturally when the fifth dimension is warped appropriately and the gauge hierarchy problem is understood without suffering from a fine-tuning problem like the SM.

Although it is sufficient to explain the gauge hierarchy problem when only the graviton propagates into the extra dimension, an extension of the RS model, where (some of) the SM particles also propagate into the bulk, has been studied from phenomenological point of view. A generic consequence of such an extension of the RS model is that there are KK excitations of the SM particles. It is, therefore, important to investigate possibilities of production and decay of KK particles at the LHC as a direct test of the model.

\section{Our scenario}
\label{sec-3}

In our scenario, we assume that the SM Higgs is located at the visible brane, while the other SM fields and gravity are present in the five-dimensional bulk. We are interested in case that the third generation of quarks couples to the KK gluons strongly, compared to the four-dimensional QCD coupling. Under this setup, we study the process $pp \rightarrow g_{KK} \rightarrow b\bar{b}$, where $g_{KK}$ is the first excitation mode of the KK gluon. 

We consider the following scenarios with various values of couplings between the KK mode of gluon and quarks:
\begin{eqnarray}
&&{g_{Q_3}^{(1)} \over g_4}={g_{t}^{(1)} \over g_4}={g_{b}^{(1)} \over g_4}=4, \quad {g_{\rm light}^{(1)} \over g_4}=0, \label{couplings1}\\
&&{g_{Q_3}^{(1)} \over g_4}=1, \quad {g_{t}^{(1)} \over g_4}={g_{b}^{(1)} \over g_4}=4, \quad {g_{\rm light}^{(1)} \over g_4}=0, \label{couplings2}\\
&&{g_{Q_3}^{(1)} \over g_4}=1, \quad {g_{t}^{(1)} \over g_4}=4,\quad {g_{b}^{(1)} \over g_4}={g_{\rm light}^{(1)} \over g_4}=0, \label{couplings3}
\end{eqnarray}
where $Q_3$ is the third generation of the left-handed quark, $t,b$ are
the right-handed top and bottom quarks and ``light'' means the quarks of
the first two generations. In (\ref{couplings1}), couplings of all the
quarks of the third generation to the KK gluon is strong while the
coupling between the KK gluon and the light quarks is vanishing. 
In (\ref{couplings2}),  the KK gluon
strongly couples to the right-handed quarks only. The coupling to the
left-handed quark is comparable to the QCD coupling $g_4$. 
In (\ref{couplings3}), the difference from
(\ref{couplings2}) is to take the KK gluon coupling to the right-handed
bottom quark to be zero.

\section{Simulation of the model}

We simulated signal $pp \rightarrow g_{KK} \rightarrow b\bar{b}$ and possible background processes, initial/final state radiation, hadronization and decays using Pythia 8.160 \cite{Pythia6,Pythia8}, the Monte-Carlo generator. All samples were generated for $pp$ collisions at $\sqrt{s} = 14$ TeV. 

For the simulation of the effects of a detector, we used Delphes 1.9 \cite{Delphes}, a framework for a fast simulation of a generic collider experiment. For the jets reconstruction, Delphes uses FastJet tool \cite{FastJet1,FastJet2}. In our simulations, we used the data file with standard settings for the ATLAS detector, provided by the tool. We used the $k_{T}$ algorithm \cite{kT_algorithm} with a cone radius parameter $R$ = 0.7. The $b$-tagging efficiency is assumed to be 40\%, independently on a transverse momentum and pseudorapidity of a jet. A fake rate of a $b$-tagging algorithm is assumed to be 10\% for $c$-jets and 1\% for light and gluon jets.

\section{Selection criteria}

We assume this set of event selection criteria:
\begin{enumerate}
  \item The event must have exactly 2 $b$-tagged jets with the transverse momentum $p_T > 100$ GeV, the pseudorapidity $|\eta| < 2.5$ and invariant mass $M_{b\bar{b}} > M_{b\bar{b}}^{min}$.
  \item The event must have no other jet with $p_T > 20$ GeV, $|\eta| < 4.9$.
  \item The event must have no electron or muon with $p_T > 10$ GeV, $|\eta| < 2.5$.
  \item The reconstructed transverse missing energy of the event $E_{T}^{miss} < 50$ GeV.
\end{enumerate}
The criterion no. 1 for sufficiently high $M_{b\bar{b}}^{min}$
effectively suppresses the QCD background processes 
. The criterion no. 2 suppresses a top-antitop pair production in both the QCD and the RS model, with the subsequent decay of the top quarks to jets. The criteria no. 3 and 4 effectively suppress other decay channels of a top quark decay. 

\section{Numerical results}

We simulated the signal process for various values of couplings and for the masses of KK gluon between 1 TeV and 1.5 TeV. All presented results are scaled to the luminosity of 10 fb$^{-1}$.

\begin{figure*}
\begin{center}
\begin{eqnarray*}
 \begin{array}{cc}
  \epsfxsize=6.5cm
  \hspace{1cm}
  \epsfbox{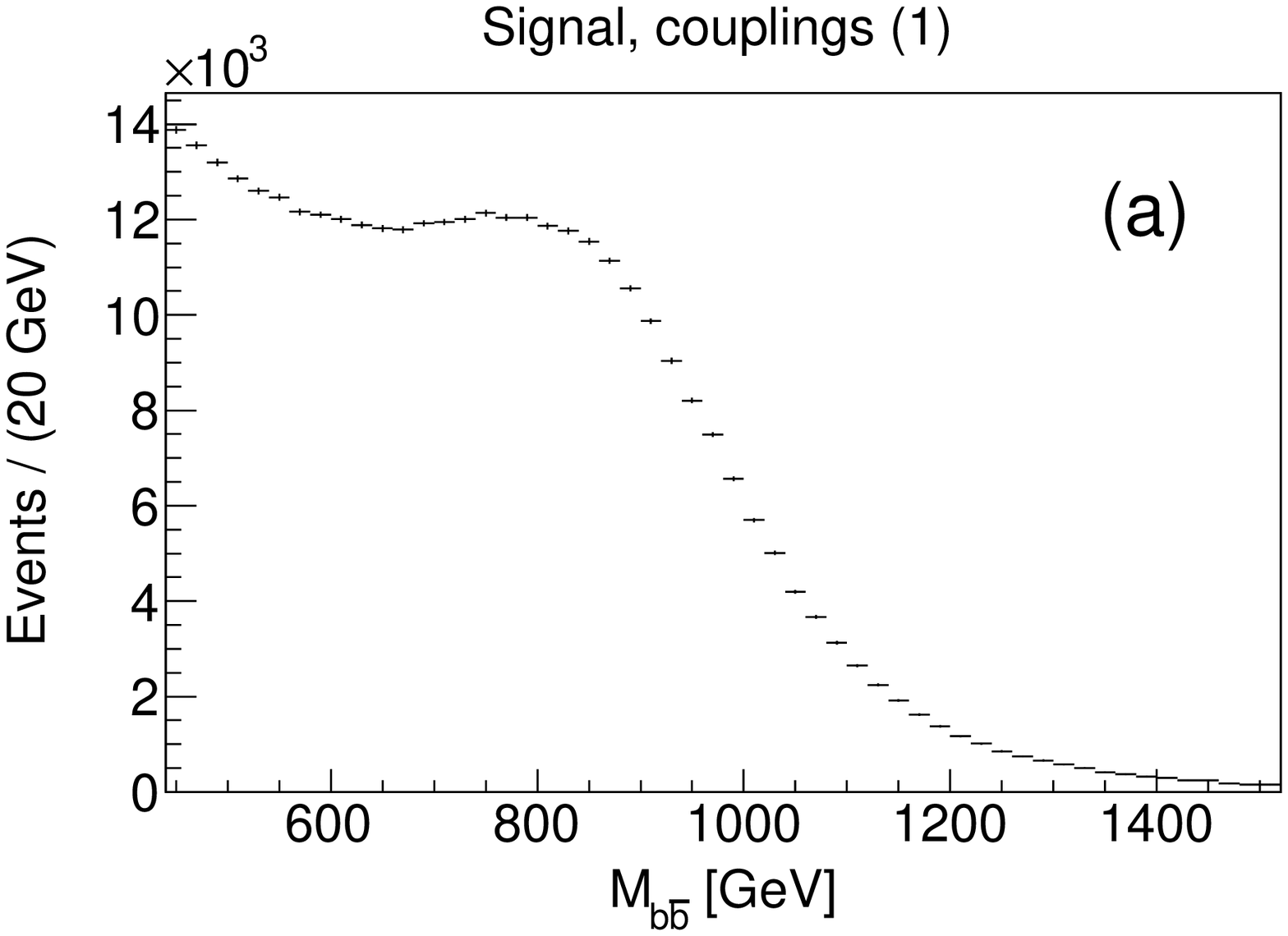}
  &
  \epsfxsize=6.5cm
  \epsfbox{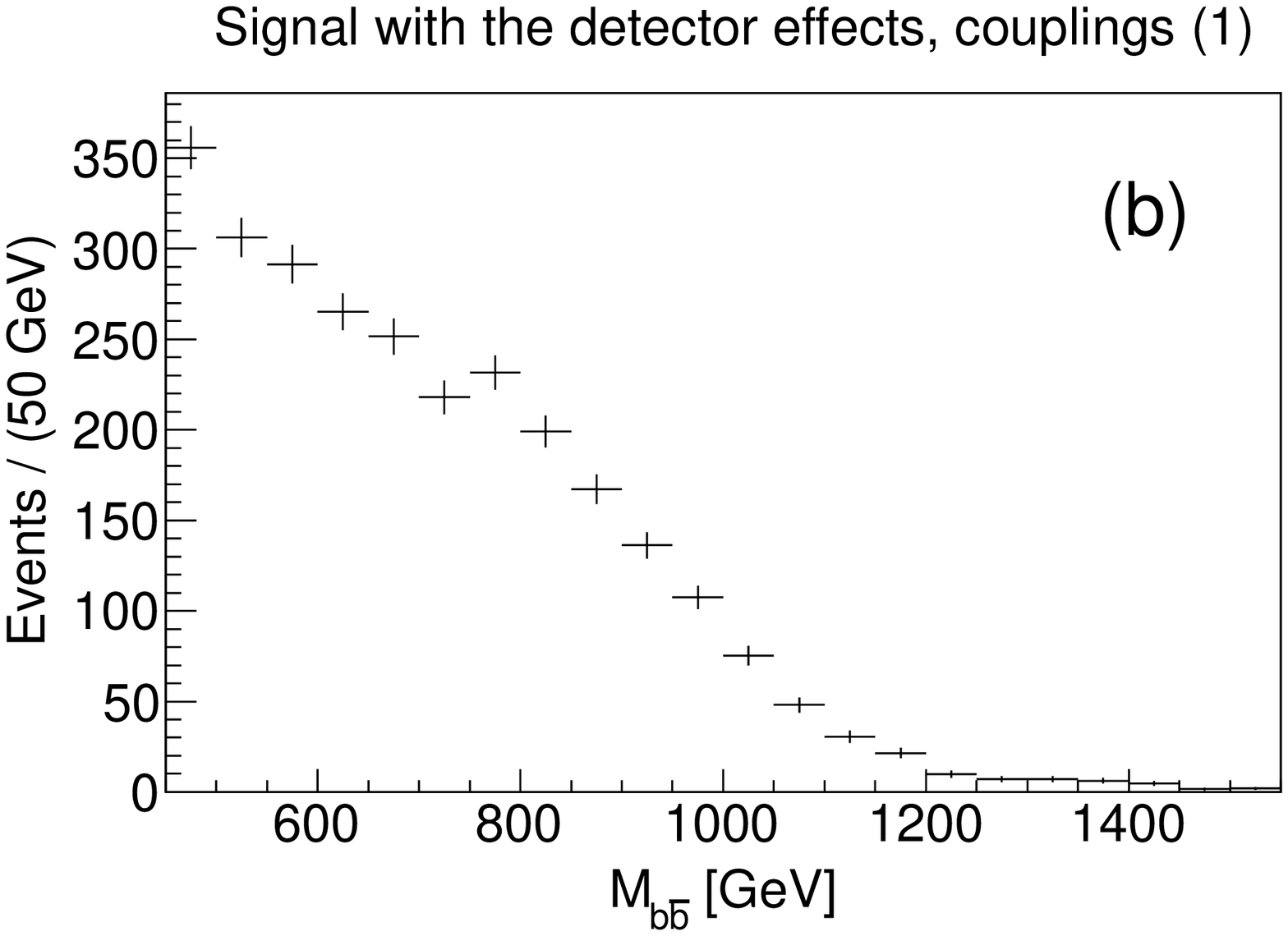} \\
  \epsfxsize=6.5cm
  \hspace{1cm}
  \epsfbox{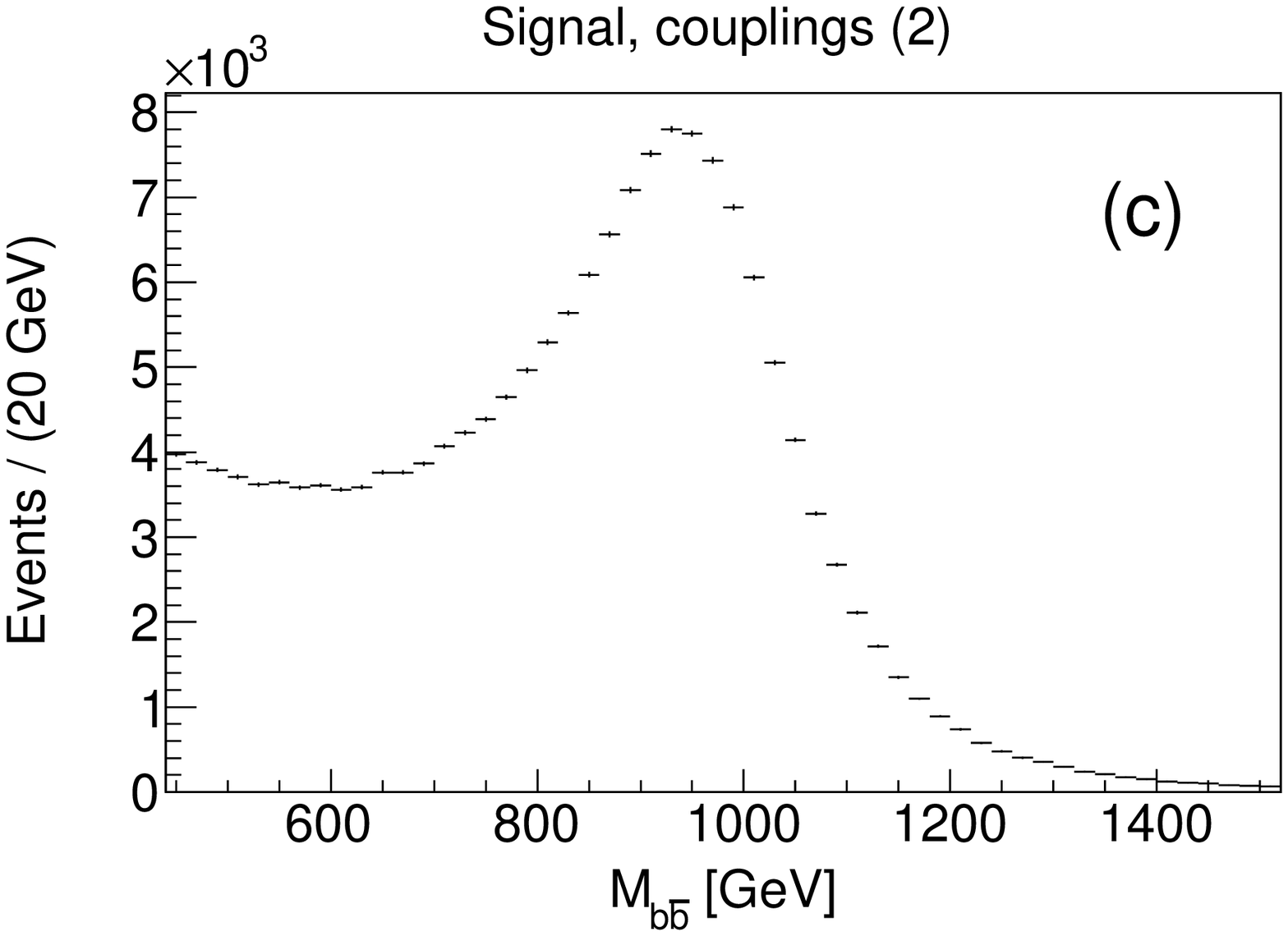}
  &
  \epsfxsize=6.5cm
  \epsfbox{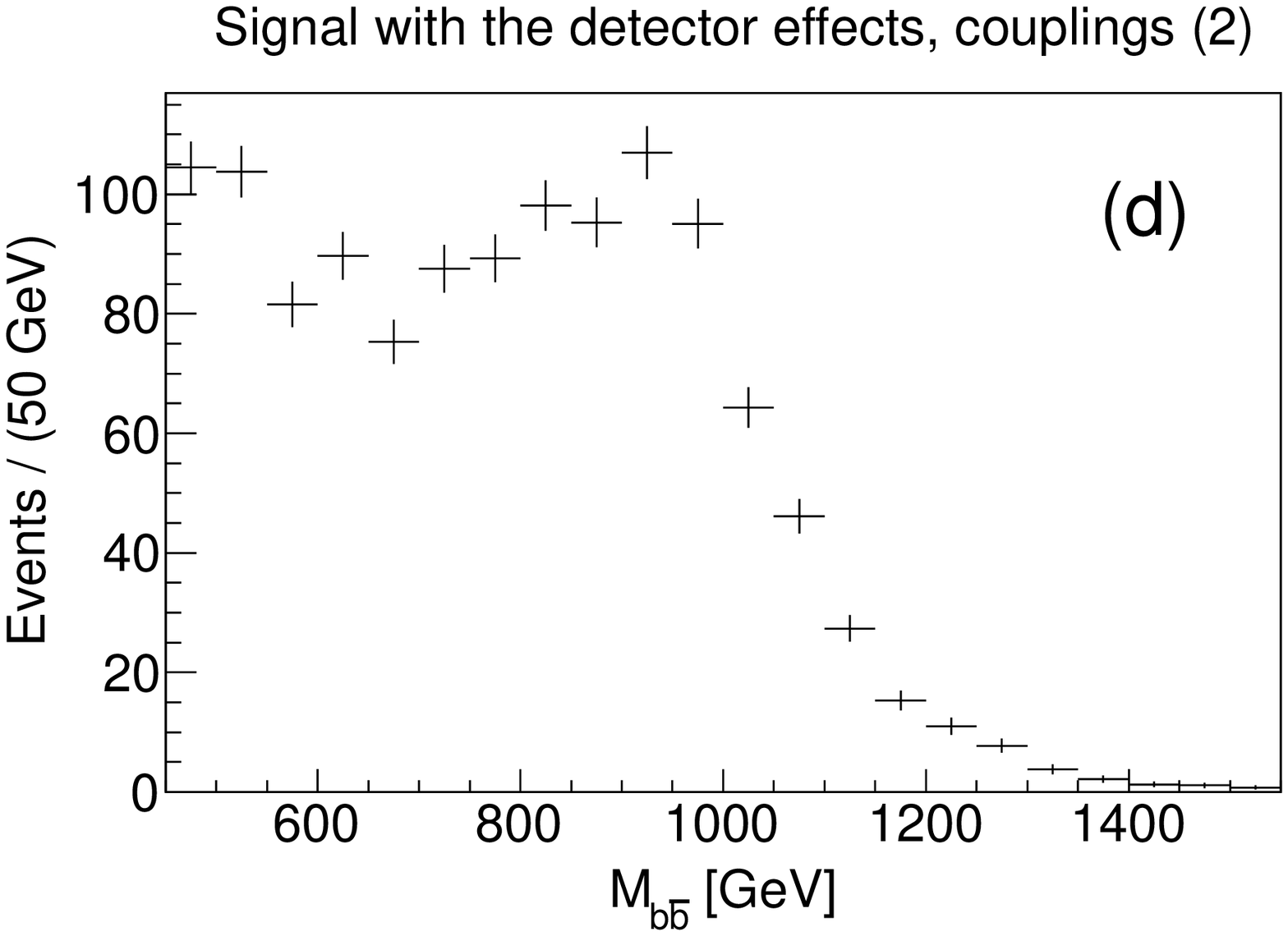} \\
  \epsfxsize=6.5cm
  \hspace{1cm}
  \epsfbox{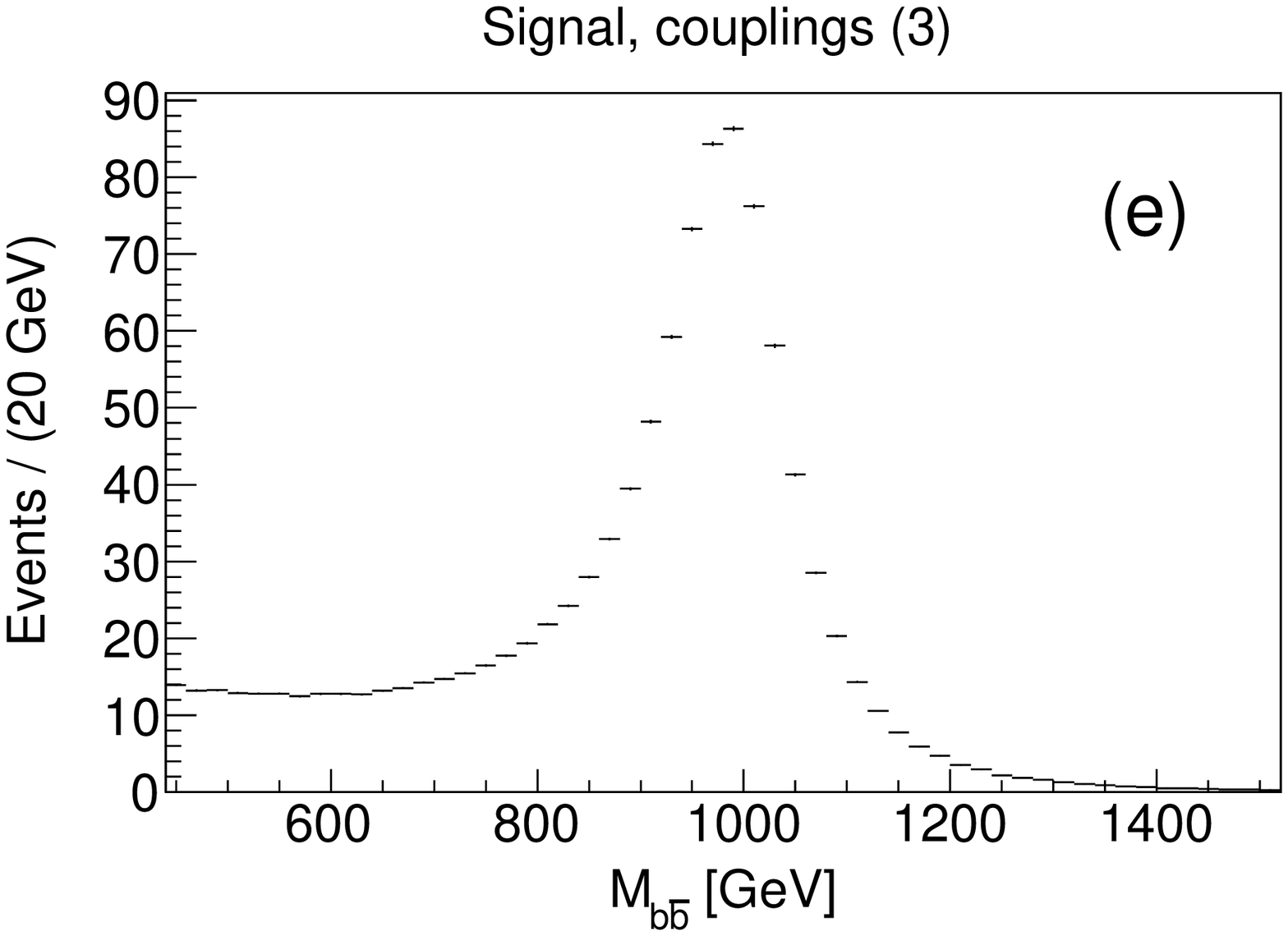}
  & 
  \epsfxsize=6.5cm
  \epsfbox{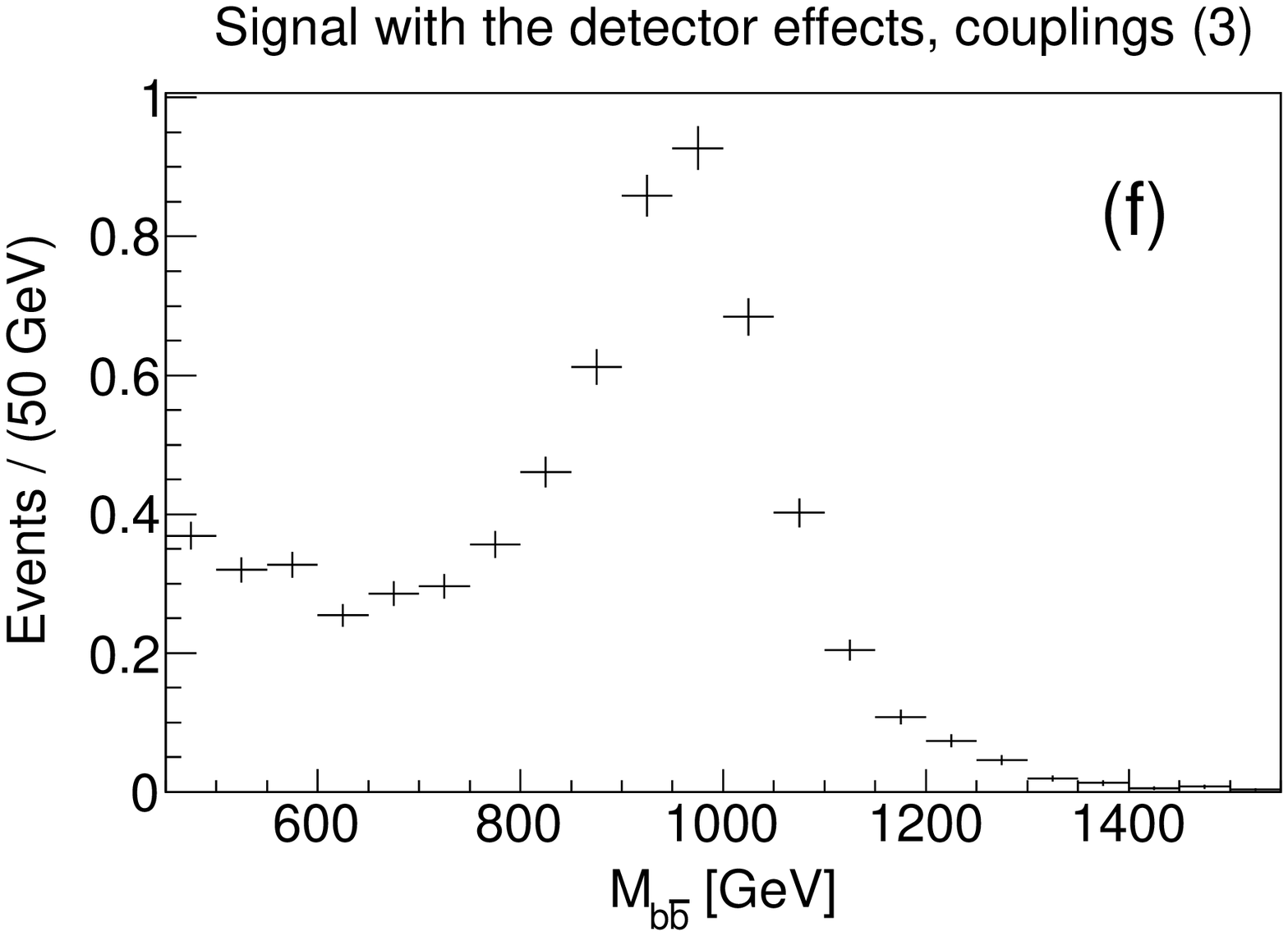} \\
 \end{array}
\end{eqnarray*}
\caption{
The invariant mass distribution of the $b\bar{b}$ pairs for the signal 
 process $pp \rightarrow g_{KK} \rightarrow b\bar{b}$ without 
 ((a), (c), and (e)) and with ((b), (d), and (f)) the simulated effects 
 of the ATLAS detector and the selection criteria (with 
 $M_{b\bar{b}}^{min}$ = 450 GeV). $M_{g_{KK}} = 1$~TeV was assumed and 
 three scenarios with couplings (\ref{couplings1})--(\ref{couplings3}) 
 were studied (marked as (1), (2), and (3), in the figure). The number 
 of events in the histogram is scaled to the integrated luminosity of 10 
 fb$^{-1}$ for $pp$ collisions at $\sqrt{s}=14$ TeV.  
}
  \label{signal}
\end{center}
\end{figure*}

In the 
 Fig. \ref{signal}, distributions of a $b\bar{b}$ invariant mass without
 and with the simulated detector effects and the selection criteria for
 $M_{g_{KK}} = 1$~TeV and couplings
 (\ref{couplings1})--(\ref{couplings3}) are presented.

As a signature of new physics, we use the number of selected events. We estimated number of expected observed signal and background events and the statistical significance $S/\sqrt{B}$. The results of the simulations are presented in the Table \ref{TabSignificance}. In the presented results, we use the value of $M_{b\bar{b}}^{min}$, for which the statistical significance $S/\sqrt{B}$ is maximal.

\begin{table}
\centering
\caption{The statistical significance $S/\sqrt{B}$ of our model for various values of $M_{g_{KK}}$ and couplings estimated for 10 fb$^{-1}$. 
}
\label{TabSignificance}
\begin{tabular}{c c c c c c}
\hline
\multirow{2}{*}{$\frac{g^{(1)}_{\rm light}}{g_4}$} & \multirow{2}{*}{$\frac{g^{(1)}_{Q_3}}{g_4}$} & \multirow{2}{*}{$\frac{g^{(1)}_b}{g_4}$} & \multirow{2}{*}{$\frac{g^{(1)}_t}{g_4}$} & $M_{g_{KK}}$ & $S/\sqrt{B}$ \\
                              &                          &                       &                       & [TeV]                      & for 10 fb$^{-1}$ \\ 
\hline
\hline
\multirow{5}{*}{0}            & \multirow{5}{*}{4}       & \multirow{5}{*}{4}    & \multirow{5}{*}{4}    & 1.0                          & $7.3 \pm 0.5$  \\
                              &                          &                       &                       & 1.1                          & $5.0 \pm 0.4$  \\
                              &                          &                       &                       & 1.2                          & $3.4 \pm 0.3$  \\ 
                              &                          &                       &                       & 1.3                          & $2.3 \pm 0.2$  \\ 
                              &                          &                       &                       & 1.4                          & $1.7 \pm 0.2$  \\ 
                              &                          &                       &                       & 1.5                          & $1.2 \pm 0.2$  \\ 
\hline
\multirow{5}{*}{0}            & \multirow{5}{*}{1}       & \multirow{5}{*}{4}    & \multirow{5}{*}{4}    & 1.0                          & $4.4 \pm 0.3$  \\
                              &                          &                       &                       & 1.1                          & $3.0 \pm 0.4$  \\
                              &                          &                       &                       & 1.2                          & $2.0 \pm 0.3$  \\ 
                              &                          &                       &                       & 1.3                          & $1.4 \pm 0.2$  \\ 
                              &                          &                       &                       & 1.4                          & $0.9 \pm 0.5$  \\ 
                              &                          &                       &                       & 1.5                          & $0.9 \pm 0.4$  \\ 
\hline
\multirow{1}{*}{0}            & \multirow{1}{*}{1}       & \multirow{1}{*}{0}    & \multirow{1}{*}{4}    & 1.0                          & $0.033 \pm 0.003$ \\
\hline
\end{tabular}
\end{table}

As expected, the deviation from the SM is strongly dependent on the coupling of a right-handed $b$-quark to a KK gluon. For the first and second set of couplings, the effects of KK gluons could be observable. Due to the extremely low cross-section of the signal process, for the third set of couplings the effects of KK gluons are unobservable.

\section{Conclusions}

We studied the possibility of observation of effects of the first excitation of a KK gluon, predicted by the extension of the RS model. In our work, we aimed on the final states with two $b$ jets. We prepared appropriate Monte-Carlo simulations of the signal and background processes for the $pp$ collisions with the energy $\sqrt{s} = 14$~TeV at the LHC, simulated the effects of the ATLAS detector and the selection criteria. As a signature of new physics, we used the number of selected events. We studied three scenarios (\ref{couplings1})--(\ref{couplings3}) with various couplings of a KK gluon to $b$ and $t$ quarks. We estimated the significance $S/\sqrt{B}$ of our model. 
For the integrated luminosity of 10~fb$^{-1}$, the deviation from the
SM could be observable with the significance of several sigmas for the mass of a KK gluon up to 1.5~TeV and scenarios (\ref{couplings1}) and (\ref{couplings2}). The effects of a KK gluon in the scenario (\ref{couplings3}) with vanishing coupling of a KK gluon to a right-handed $b$ quark will not be observable.

\section{Acknowledgements}

The work of M.A. and K.S. is supported in part by the Research Program
MSM6840770029, by the project of International Cooperation ATLAS-CERN 
and by the project LH11106 of the Ministry of Education, Youth and Sports 
of the Czech Republic. 
The work of G.C.C is supported in part by Grants-in-Aid for Scientific 
Research from the Ministry of
Education, Culture, Sports, Science and Technology (No.24104502) and 
from the Japan Society for the Promotion of Science (No.21244036).

%
%
%

\end{document}